# Predicting Fracture Energies and Crack-Tip Fields of Soft Tough Materials

*Teng Zhang[1,a), Shaoting Lin[1,a), Hyunwoo Yuk[1], Xuanhe Zhao[1,2, b)]*

[1.] *Soft Active Materials Laboratory, Department of Mechanical Engineering, Massachusetts Institute of Technology, Cambridge, MA 02139, USA;* [2.] *Department of Civil and Environmental Engineering, Massachusetts Institute of Technology, Cambridge, MA 02139, USA*

a) These authors contribute equally to this work

b) To whom correspondence should be addressed. Email: zhaox@mit.edu

## Abstract

Soft materials including elastomers and gels are pervasive in biological systems and technological applications. Whereas it is known that intrinsic fracture energies of soft materials are relatively low, how the intrinsic fracture energy cooperates with mechanical dissipation in process zone to give high fracture toughness of soft materials is not well understood. In addition, it is still challenging to predict fracture energies and crack-tip strain fields of soft tough materials. Here, we report a scaling theory that accounts for synergistic effects of intrinsic fracture energies and dissipation on the toughening of soft materials. We then develop a coupled cohesive-zone and Mullins-effect model capable of quantitatively predicting fracture energies of soft tough materials and strain fields around crack tips in soft materials under large deformation. The theory and model are quantitatively validated by experiments on fracture of soft tough materials under large deformations. We further provide a general toughening diagram that can guide the design of new soft tough materials.

**Predicting Fracture Energies and Crack-Tip Fields of Soft Tough Materials**

**Introduction.**  Except bones and teeth, most parts of animal bodies consist of soft materials – elastomers and hydrogels with relatively low rigidity and high deformability compared to hard materials such as steel and ceramics. Biological soft materials such as cartilage, muscle, skin and tendon usually need to maintain high toughness, which is critical for survival and well-being of animals under various internal and external loads [1]. Soft materials also promise broad technological applications in areas as diverse as soft machines and robots [2-4], artificial tissues and organs [5], non-conventional electronics [6,7], and microfluics and optics [8,9]. In these applications, high toughness of the materials is usually required for reliability and robust function of the systems.

Owing to their scientific and technological importance, various soft tough materials have been developed in recent decades [10-13]. The intrinsic fracture energy of soft materials – i.e., the energy required to fracture a layer of polymer chains in front of the crack [14] – is relatively low; and it is qualitatively known that the toughening of soft materials generally relies on mechanical dissipation in process zones around cracks [14-19]. However, it is still not well understood how the intrinsic fracture energy and mechanical dissipation cooperate synergistically to give rise to high fracture toughness of soft materials. Furthermore, physical models that can predict the fracture energy and crack-tip strain fields of soft materials are of imminent importance to the design of new soft tough materials, but such predictive models still do not exist.

Here, we report a scaling law and a continuum model that quantitatively accounts for the synergistic contributions of intrinsic fracture energies and dissipations to the total fracture energies of soft materials. We characterize the essential physical features of intrinsic fracture energy and dissipation using the cohesive-zone model and Mullins-effect model, respectively, implemented in finite-element software, ABAQUS. Our calculation shows that the total fracture energy of soft material scales linearly with its intrinsic fracture energy, while the effect of dissipation manifests as a scaling pre-factor that can be much higher than one. To validate the theory and model, we measure the stress-strain hysteresis and intrinsic fracture energies of polyacrylamide-alginate (PAAm-alginate) hydrogels of different compositions, which represent soft tough materials with different properties [12,20]. Using the material parameters measured independently, our model can quantitatively predict the fracture energies of different soft materials as well as strain fields and crack propagations in them. Based on the model, we further calculate a toughening diagram that can guide the design of new soft tough materials.

**Scaling analysis –** Let's consider a notched soft material undergoing the pure-shear test to measure its fracture energy, as illustrated in **Fig. 1a** [21]. Crack propagation in the soft material requires the scission of at least a layer of polymer chains. The required mechanical energy for chain scission divided by the area of crack surface at undeformed state gives the intrinsic fracture energy, $\Gamma_0$. In addition, material elements in a process zone around the crack will also be deformed and undeformed as the crack propagates. If mechanical energy is dissipated during this process, the dissipated energy divided by the area of crack surface at undeformed state further contributes to the total fracture energy by, $\Gamma_D$ (**Fig.1b**). Therefore, the total fracture energy of a soft material can be expressed as

$$\Gamma = \Gamma_0 + \Gamma_D ,$$  (1)

where $\Gamma_D = \overline{U}_D l_D$ and $\overline{U}_D$ is the mechanical energy dissipated per the volume of the process zone, and $l_D$ the height of the process zone in the soft material at undeformed state. Since material elements in the process zone mainly undergo the pure-shear deformation, we further have $\overline{U}_D \propto U_D(S = S_{max})$, where $U_D$ is the mechanical energy dissipated per unit volume of the soft material under pure-shear deformation, and $S_{max}$ is the maximum nominal stress that can be achieved in the material under pure-shear deformation (**Fig. 1b**).

Before a crack propagates in a material, a cohesive zone with finite length will be formed along the crack path [22]. For a soft material, represented by the neo-Hookean model, the length of the cohesive zone scales as $l_{cohesive} \propto \mu \Gamma / S_{max}^2$, where $\mu$ is the shear modulus of the materials, $\Gamma$ the fracture toughness and $S_{max}$ the maximum nominal stress (See supplementary materials and Ref [23-25] for details). In addition, since the cohesive zone is encapsulated in the process zone around the crack, the size of the process zone scales with the length of the cohesive zone, i.e.,

$$l_D \propto l_{cohesive} \propto \Gamma / U_{max} \qquad (2)$$

where $U_{max} \propto S_{max}^2 / \mu$ is the maximum mechanical work done on the material under pure-shear deformation. A combination of Eq. [1] and [2] leads to a governing equation for the total fracture energy of soft tough materials [26-28],

$$\Gamma = \frac{\Gamma_0}{1 - \alpha \cdot h_{max}} \qquad (3)$$

where $h_{max} = U_D(S = S_{max})/U_{max}$ is the ratio between the maximum dissipation and maximum mechanical work done on the material, and $0 \le \alpha \le 1$ is a dimensionless number depending on the stress-strain hysteresis of the material deformed to different levels of stresses.

**Coupled cohesive-zone and Mullins-effect model** – Next we develop a continuum model that can predict the fracture energy of soft tough materials, using material parameters measured independently. The continuum model needs to quantitatively capture the synergistic contributions of intrinsic fracture energy and mechanical dissipation in process zone to the total fracture energy. In order to model the intrinsic fracture energy of soft materials, we adopt a triangle cohesive-zone model governed by the maximum nominal stress ($S_{\max}$) and maximum nominal separation ($\delta_{\max}$) on the expected crack path (**Fig. 1c**) [23]. The damage initiation of the cohesive layer follows the quadratic nominal stress criterion

$$\left\{ \frac{t_n}{S_{\max}} \right\}^2 + \left\{ \frac{t_s}{S_{\max}} \right\}^2 = 1 \tag{4}$$

where $t_{(\bullet)}$ represents the nominal surface tractions on the crack surface, and the subscripts $n$ and $s$ indicate normal and tangential directions, respectively. When Eq. (4) is satisfied, the cohesive layer enters into the softening regime, which is described by the linear damage evolution function illustrated in **Fig. 1c**. The cohesive-zone model prescribes the intrinsic fracture energy of the soft materials to be,

$$\Gamma_0 = 1/2 \, S_{\max} \delta_{\max} \tag{5}$$

To physically implement the cohesive-zone model, the maximum nominal stress $S_{\max}$ of the cohesive zone is taken as the measured failure stress of the material under pure-shear tension, and the maximum nominal separation $\delta_{\max}$ is calculated based on the experimentally measured intrinsic fracture energy of the material $\Gamma_0$ and Eq. (5), i.e., $\delta_{\max} = 2\Gamma_0 / S_{\max}$.

Mechanical dissipation in the process zone may arise from viscoelasticity [29], plasticity [30], and/or partial damage of the soft materials; and such dissipations manifest as hysteresis loops on stress-strain curves of the material (**Fig. 1b**). To capture the essential effect of dissipation in process zone on toughening, we model the dissipation as the Mullins effect in soft materials [31]. The Mullins effect gives hysteresis loops in the stress-stretch curves of the materials under loading-unloading cycles. We define the hysteresis ratio of the material under pure-shear deformation as

$$h = U_D/U \tag{6}$$

where $U$ and $U_D$ are the mechanical work done on and the energy dissipation in a unit volume of the soft material under pure-shear deformation, respectively. The hysteresis ratio of soft material generally increases with the deformation of (or the work done on) the material due to the accumulation of material damage and eventually reaches a maximum value, i.e., $h(S = S_{max}) = h_{max}$ (**Fig. 1d**), which is used in Eq. (3). To describe the Mullins effect in soft materials, we adopt the modified Ogden-Roxburgh model used in ABAQUS [32]. In brief, the free energy function of an incompressible material with Mullins effect can expressed as

$$W(\mathbf{F},\eta) = \eta \tilde{W}(\mathbf{F}) + \phi(\eta) \tag{6}$$

where $\mathbf{F}$ is the deformation gradient tensor, $\eta$ is a damage variable ($0 < \eta \leq 1$), $\tilde{W}$ is the free energy function of a pure elastic material without Mullins effect, and $\phi(\eta)$ is referred to as the damage function. The damage function and damage variable in Eq. (6) can be expressed as

$$\phi(\eta) = \int_1^\eta \left[ \left(m + \beta W^m\right) \mathrm{erf}^{-1}\left(r(1-\eta)\right) - W^m \right] d\eta \tag{7a}$$

$$\eta = 1 - \frac{1}{r} \mathrm{erf}\left[ \left(W^m - \tilde{W}\right) / \left(m + \beta W^m\right) \right] \tag{7b}$$

where $W^m$ denotes the maximum strain energy density of the material before unloading, erf is the error function, $\beta$ is a positive number to avoid overly stiff response at the initiation of unloading from relatively large stretch levels, and $r$ and $m$ are constants that characterize the damage properties of the material. Throughout the calculations, we set $\beta$=0.1 for numerical stabilization. The parameter $r$ in Eq. (7) indicates the maximum extent of the material damage related to the virgin state [31], which therefore determines the maximum hysteresis ratio of the material under pure-shear deformation, $h_{max}$. **Figure 2a** gives the relation between $h_{max}$ and $r$ for a neo-Hookean material with Mullins effect under pure-shear deformation. It is evident that $h_{max}$ is a monotonic decreasing function of $r$. The parameter $m$ in Eq. (7) represents a critical energy scale that acts as a threshold for activating significant dissipation in the material. **Figure 2b** shows the calculated hysteresis ratio $h$ as a function of $U/U_{max}$ for different values of $m/U_{max}$ for a neo-Hookean material with Mullins effect under pure-shear deformation. If $U/U_{max} < m/U_{max}$ for a material under pure-shear deformation, the hysteresis ratio $h$ is generally much smaller than $h_{max}$, which means that the deformation of (or the work done on) the material is not sufficient to induce significant dissipation. If $U/U_{max} >> m/U_{max}$, the hysteresis ratio $h$ can reach a value close to $h_{max}$, which means that significant dissipation has been activated. (See **Fig. 1d** for schematics and **Fig. 2c** for calculation.) Therefore, the parameter $m/U_{max}$ indicates the speed of $h$ increasing from 0 to $h_{max}$ as a function of $U/U_{max}$; a smaller value of $m/U_{max}$ gives a faster transition to $h_{max}$.

To physically implement the Mullins-effect model, the free energy function $\tilde{W}(\mathbf{F})$ in Eq. (6) can be obtained by fitting a hyperelastic model to the stress-stretch curve of soft material under monotonic loading. The parameters $r$ and $m$ in Eq. (7) can be obtained from multiple stress-stretch hysteresis of the soft material deformed to different stretches.

In order to calculate the total fracture energy and crack-tip strain field of soft material, the pure-shear test is simulated in the coupled cohesive-zone and Mullins-effect model [21] (**Fig. S1 and Supplementary materials for details**). In brief, two identical pieces of a soft material are clamped along their long edges with rigid plates. A notch is introduced into the first sample, which is then gradually pulled to a stretch of $\lambda_c$ times of its undeformed length until a crack steadily propagates from the notch (**Fig. S1a**). Thereafter, the second sample without notch is uniformly stretched to the same critical stretch $\lambda_c$ with the applied stress $S$ recorded (**Fig. S1b**). The total fracture energy of the soft material can be calculated as $\Gamma = L_0 \int_1^{\lambda_c} S d\lambda$, where $L_0$ is the height of the sample shown in **Fig. S2a**.

**Theoretical and numerical results -** Next, we will use the coupled cohesive-zone and Mullins-effect model to validate the scaling for total fracture energies of soft materials. The soft material under loading is taken as a neo-Hookean material with initial shear modulus $\mu$. Based on the coupled cohesive-zone and Mullins effect model, the total fracture energy of the soft material, $\Gamma$, is mainly determined by a set of four parameters including $\mu$, $\Gamma_0$, $h_{\max}$, $S_{\max}$, and $m/U_{\max}$. [Note that $U_{\max}$ is approximately equal to $S_{\max}^2/2\mu$ for neo-Hookean materials.] We will vary these parameters independently in the model, and calculate the fracture energy of the materials following the pure-shear method described above. Without loss of generality, we use the initial shear modulus $\mu$ to normalize $S_{\max}$, $\Gamma_0$ and $\Gamma$. In **Fig. 3a**, the calculated values of $\Gamma$ are plotted as functions of $\Gamma_0$ of materials with different combinations of $S_{\max}/\mu$ and $h_{\max}$. It can be seen that $\Gamma$ is linearly scaled with $\Gamma_0$ in all the calculated cases. With this knowledge in mind, we next explore the enhanced ratio of the fracture energy ($\Gamma/\Gamma_0$) due to mechanical dissipation in the process zone. We calculate $\Gamma/\Gamma_0$ as functions of $h_{max}$ for various combinations of $S_{max}/\mu$ and

$m/U_{max}$. **Figure 3b** shows that the relation $\Gamma/\Gamma_0 = 1/(1 - \alpha \cdot h_{max})$ is valid for wide ranges of

$S_{max}/\mu$ (i.e., from 2 to 6) and $m/U_{max}$ (i.e., 0.005 to 0.1). In addition, it can be seen that the

calculated values of $\Gamma/\Gamma_0$ from models with different $S_{max}/\mu$ but the same $m/U_{max}$ are

approximately the same. This means the parameter $\alpha$ in Eq. (4) mainly depends on the

normalized critical energy scale, $m/U_{max}$. In **Fig. 3c**, we summarize the calculated parameter $\alpha$

as a function of $m/U_{max}$ for different values of $S_{max}/\mu$. It is evident that $\alpha$ does not depend on

$S_{max}/\mu$, which is consistent with the result in **Fig. 3b**. In addition, $\alpha$ is a monotonic decreasing

function of $m/U_{max}$. This trend can be qualitatively understood as follow. When the normalized

work done on a material element in the process zone exceeds a critical value $m/U_{max}$, the

element begins to dissipate mechanical energy significantly. Therefore, for materials with

otherwise the same properties, a lower value of $m/U_{max}$ gives more dissipation in the process

zone and thus a higher enhancement of the fracture energy, i.e., higher value of $\Gamma/\Gamma_0$. Based on

the models' results (**Fig. 3c**), we further fit $\alpha$ as a function of $m/U_{max}$ for neo-Hookean

materials as

$$\alpha \approx 0.33 + \frac{0.034}{m/U_{max} + 0.045}. \tag{8}$$

Based on Eq. (3) and (8), we summarize the toughness enhancement of soft materials,

$\Gamma/\Gamma_0$, as a function of the maximum hysteresis ratio $h_{max}$ and the normalized critical energy

scale for significant dissipation $m/U_{max}$ in **Fig. 4**. The results reveal three critical factors in toughening of soft materials: (1) high intrinsic fracture energy (*i.e.*, high $\Gamma_0$), (2) high value of maximum hysteresis ratio (*i.e.,* high $h_{max}$), and (3) quick transition to the maximum hysteresis (*i.e.,* low $m/U_{max}$). These findings are consistent with the underlying physical mechanisms for the design of soft tough materials, such as large amounts of long stretchy polymer chains for high intrinsic fracture energy, sacrificial bonds for high energy dissipation, and high strechability for a quick transition from zero to maximum hysteresis ratio. Our theoretical models can quantify the contributions from each factor, and therefore provide quantitative guidelines for the design of future soft tough materials.

**Experimental validation -** To validate the proposed theory and model, we take the interpenetrating-network hydrogel of polyacrylamide-alginate as a model soft tough material. (Details of the material synthesis are given in the supplementary materials.) We measure stress-stretch curves and various hysteresis ratios of the hydrogel under pure-shear deformation up to the maximum stress $S_{max}$, and then implement the measured data into the modified Ogden-Roxburgh model in ABAQUS. As shown in **Fig. 5a**, the pure elastic deformation of the hydrogel (sample 1) can be well described by the Ogden hyperelastic model [33]. In **Fig. 5b**, we compare the measured hysteresis ratio of the hydrogel under different deformation (i.e., different $U/U_{max}$) with the model's prediction, validating that the Ogden-Roxburgh model can accurately characterize the dissipative property of the hydrogel. In order to measure the intrinsic fracture energy of the hydrogel, we pre-deform the hydrogel samples to a level of stress approximately $S_{max}$ for multiple cycles to deplete the dissipative capacity of the samples [12,20]. Thereafter, the pure-shear test is used to measure the stress-stretch hysteresis and fracture energy of the pre-deformed sample. There is almost no stress-stretch hysteresis of the sample pre-deformed to $S_{max}$ [20], indicating negligible mechanical dissipation of the sample. In addition, the measured fracture energy as a function of pre-deformation indeed reaches an asymptote (**Fig. 5c**), which gives the intrinsic fracture energy of the hydrogel without the effect of dissipation in the process zone. The measured intrinsic energy is then implemented through the cohesive-zone model in ABAQUS.

Now that the material parameters of the hydrogel have been independently measured and implemented in the continuum model, we will perform the pure-shear tests on samples both in experiments and in the model to obtain the fracture energies of the hydrogel. In **Fig. 5d**, we compare the force-displacement curves of the notched sample from experiment and calculation, and find that the theoretically predicted curve and critical point for steady-state crack propagation are in good agreement with experimental results. We further use digital image correlation (DIC) method (see details in **Fig. S4**) to measure the strain field around the notch in the sample under pure-shear deformation. As shown in **Fig. 5e-f** and the supplementary movie, the strain fields around the notch predicted by the model are consistent with the measured results by DIC. One of the advantages of our numerical simulations is to visualize the distribution of the exact amount of the energy dissipation in the materials, while it is extremely challenging to obtain such quantitative data experimentally [13,34]. From **Fig. 5g**, it can be seen that a region of significant dissipation indeed encapsulates the crack tip, and the area of the region gradually increases with the external load until crack propagation. To further validate the predictive capability of the model, we fabricate another polyacrylamide-alginate hydrogel with a different composition and therefore different mechanical properties, referred to as sample 2. We then perform the same pure-shear experiment and simulation on sample 2, and show that the experiment and simulation results agree well with each other (see details in **Fig. S5**).

**Conclusion -** In this paper, we propose a scaling law that accounts for synergistic effects of intrinsic fracture energies and mechanical dissipations on the toughening of soft materials. We then develop a coupled cohesive-zone and Mullins effect model to quantitatively predict the fracture energies and crack-tip strain fields in soft tough materials, using material parameters measured independently. The theory and the model show that the toughening of a soft material relies on high intrinsic fracture energy of the material, high value of maximum hysteresis ratio of the material, and quick transition to the maximum hysteresis in the material under deformation. We further perform pure-shear experiments coupled with DIC on tough hydrogels to measure their fracture energies and strain fields around crack tips, and show that the experimental results match well with the model's predictions. The theory and model can provide quantitative guidance for the design of future soft tough materials.

## Acknowledgements

The authors acknowledge Prof. Lallit Anand for helpful discussions on cohesive-zone model. The work is supported by ONR (No. N00014-14-1-0528) and MIT Institute of Soldier Nanotechnology. The authors are also grateful for the support from MIT research computing resources.

# Reference

[1]     N. Simha, C. Carlson, and J. Lewis, Journal of Materials Science: Materials in Medicine **15**, 631 (2004).

[2]     S. A. Morin, R. F. Shepherd, S. W. Kwok, A. A. Stokes, A. Nemiroski, and G. M. Whitesides, Science **337**, 828 (2012).

[3]     Z. Suo, Mrs Bulletin **37**, 218 (2012).

[4]     H. Gao, X. Wang, H. Yao, S. Gorb, and E. Arzt, Mechanics of Materials **37**, 275 (2005).

[5]     K. Y. Lee and D. J. Mooney, Chemical reviews **101**, 1869 (2001).

[6]     J. A. Rogers, T. Someya, and Y. Huang, Science **327**, 1603 (2010).

[7]     C. Yu *et al.*, Advanced Materials **25**, 1541 (2013).

[8]     S. R. Quake and A. Scherer, Science **290**, 1536 (2000).

[9]     S. R. Sershen, G. A. Mensing, M. Ng, N. J. Halas, D. J. Beebe, and J. L. West, Advanced Materials **17**, 1366 (2005).

[10]    J. P. Gong, Y. Katsuyama, T. Kurokawa, and Y. Osada, Advanced Materials **15**, 1155 (2003).

[11]    K. J. Henderson, T. C. Zhou, K. J. Otim, and K. R. Shull, Macromolecules **43**, 6193 (2010).

[12]    J.-Y. Sun, X. Zhao, W. R. Illeperuma, O. Chaudhuri, K. H. Oh, D. J. Mooney, J. J. Vlassak, and Z. Suo, Nature **489**, 133 (2012).

[13]    E. Ducrot, Y. Chen, M. Bulters, R. P. Sijbesma, and C. Creton, Science **344**, 186 (2014).

[14]    G. Lake and A. Thomas, Proceedings of the Royal Society of London. Series A. Mathematical and Physical Sciences **300**, 108 (1967).

[15]    H. R. Brown, Macromolecules **40**, 3815 (2007).

[16]    R. E. Webber, C. Creton, H. R. Brown, and J. P. Gong, Macromolecules **40**, 2919 (2007).

[17]    Y. Tanaka, EPL (Europhysics Letters) **78**, 56005 (2007).

[18]    J. P. Gong, Soft Matter **6**, 2583 (2010).

[19]    X. Zhao, Soft matter **10**, 672 (2014).

[20]    S. Lin, Y. Zhou, and X. Zhao, Extreme Mechanics Letters  (2014).

[21]    R. Rivlin and A. G. Thomas, Journal of polymer Science **10**, 291 (1953).

[22]    A. Turon, C. G. Davila, P. P. Camanho, and J. Costa, Engineering fracture mechanics **74**, 1665 (2007).

[23]    C.-Y. Hui, A. Jagota, S. Bennison, and J. Londono, Proceedings of the Royal Society of London. Series A: Mathematical, Physical and Engineering Sciences **459**, 1489 (2003).

[24]    V. R. Krishnan, C. Y. Hui, and R. Long, Langmuir **24**, 14245 (2008).

[25]    R. Long and C.-Y. Hui, Extreme Mechanics Letters  (2015).

[26]    R. McMeeking and A. Evans, Journal of the American Ceramic Society **65**, 242 (1982).

[27]    B. Budiansky, J. Hutchinson, and J. Lambropoulos, International Journal of Solids and Structures **19**, 337 (1983).

[28]    A. G. Evans, Z. Ahmad, D. Gilbert, and P. Beaumont, Acta metallurgica **34**, 79 (1986).

[29]    T. Baumberger, C. Caroli, and D. Martina, Nature materials **5**, 552 (2006).

[30]    T. L. Sun, T. Kurokawa, S. Kuroda, A. B. Ihsan, T. Akasaki, K. Sato, M. A. Haque, T. Nakajima, and J. P. Gong, Nature materials **12**, 932 (2013).

[31]    R. Ogden and D. Roxburgh, Proceedings of the Royal Society of London. Series A: Mathematical, Physical and Engineering Sciences **455**, 2861 (1999).

[32]    D. Systèmes, Simulia Corp. Providence, RI, USA  (2007).

[33]    R. W. Ogden, *Non-linear elastic deformations* (Courier Dover Publications, 1997).

[34]    Q. Wang, G. R. Gossweiler, S. L. Craig, and X. Zhao, Journal of the Mechanics and Physics of Solids  (2015).

**Figures and Figure Captions**

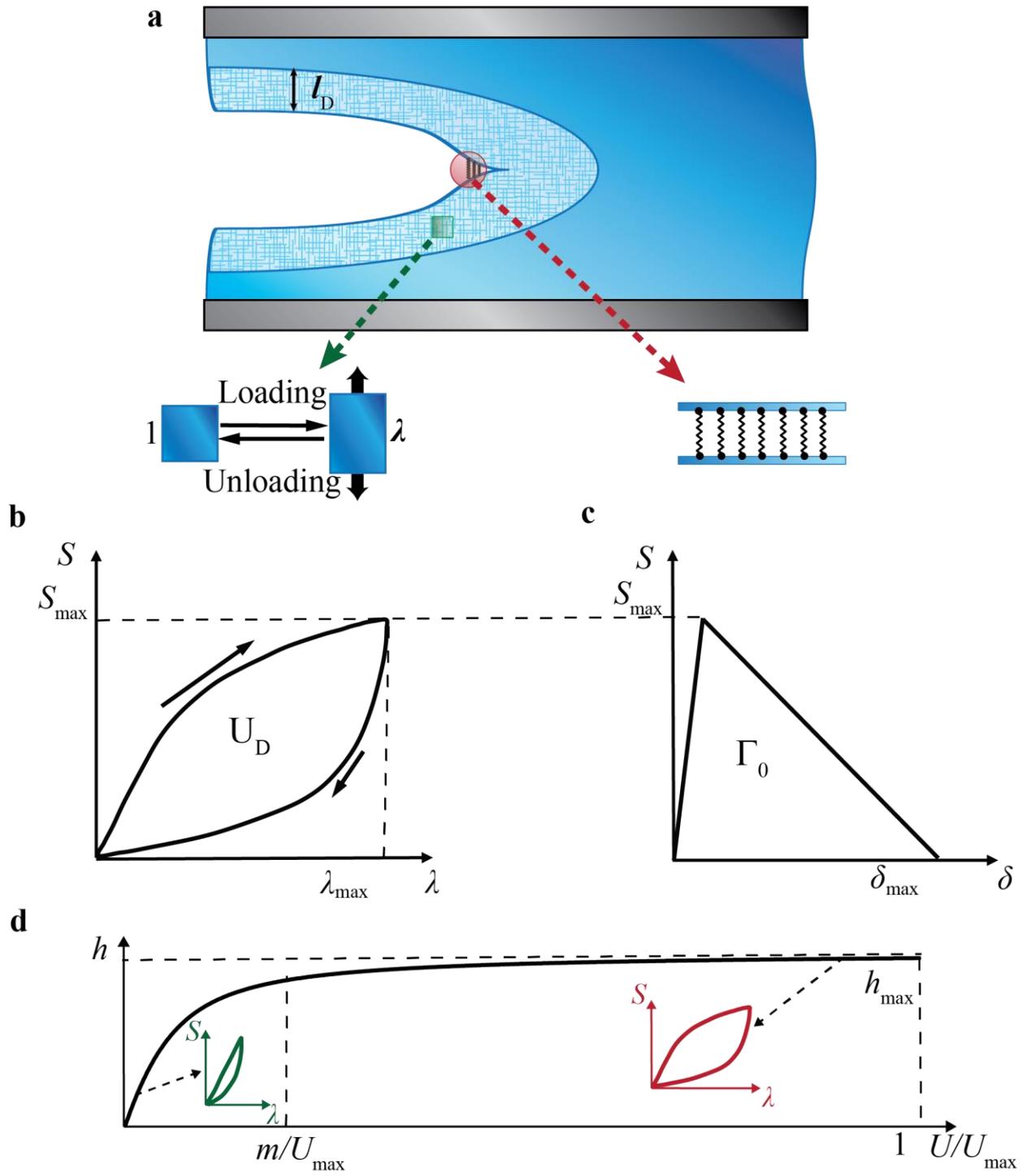

**Figure. 1**. **Schematics of the theory and model for fracture in soft tough materials.** (a) Crack propagation in a soft tough material under pure-shear test. A process zone with height $l_D$ develops in the material during crack propagation. (b) The mechanical dissipation in the process zone is characterized by the Mullins effect. A typical stress-stretch curve of the soft material under cyclic pure-shear deformation. The hysteresis loop in the curve indicates mechanical dissipation. (c) The intrinsic fracture energy of the soft material is characterized as a cohesive-zone model with triangle traction-separation law. (d) The hysteresis ratio of the soft material monotonically increases with the maximum work done to the material.

**a**

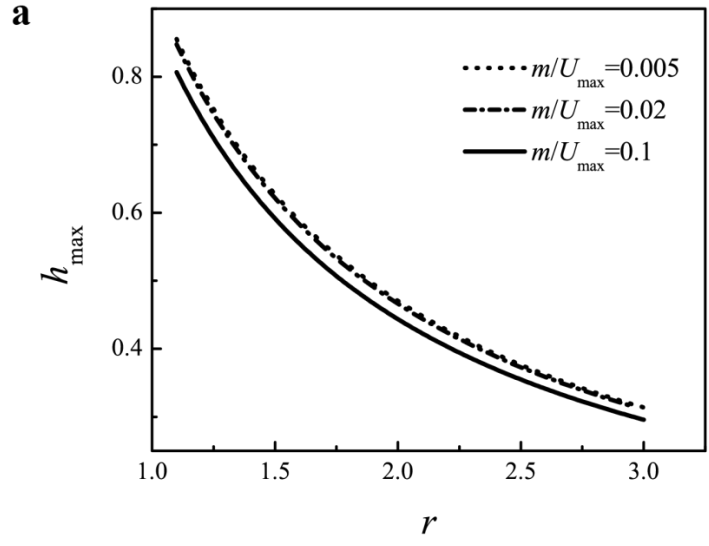

**b**

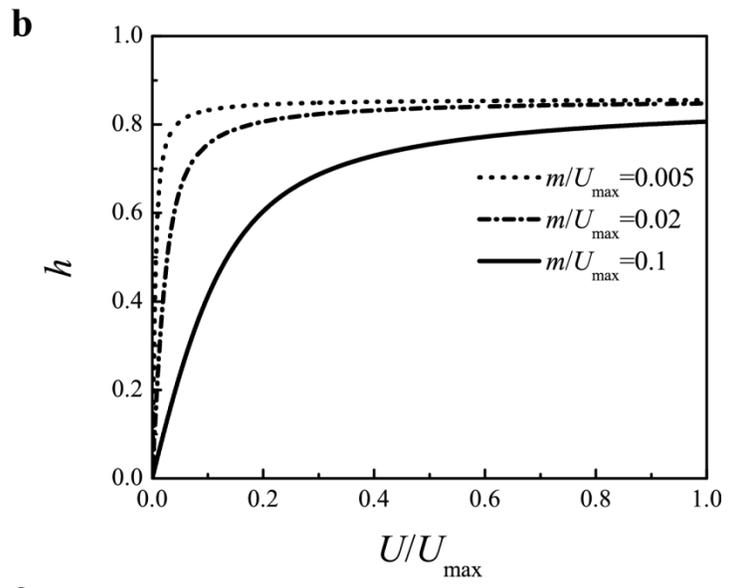

**c**

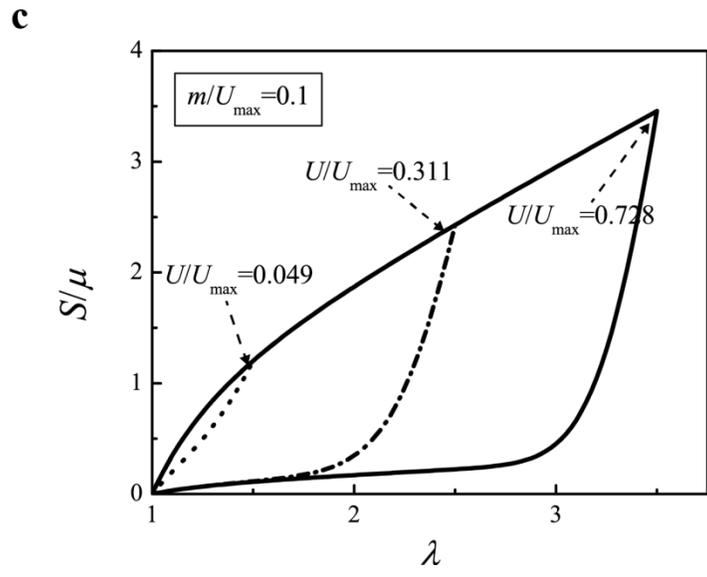

**Figure 2. The modified Ogden-Roxburgh model for Mullins effect**. (a) The relation between $h_{\max}$ and $r$ for a neo-Hookean material with Mullins effect under pure-shear deformation. (b) The calculated $h$ as a function of $U/U_{\max}$ for different values of $m/U_{\max}$ for a neo-Hookean material with Mullins effect under pure-shear deformation. (c) The calculated stress-stretch hysteresis for a neo-Hookean material with Mullins effect under pure-shear deformation.

**a**

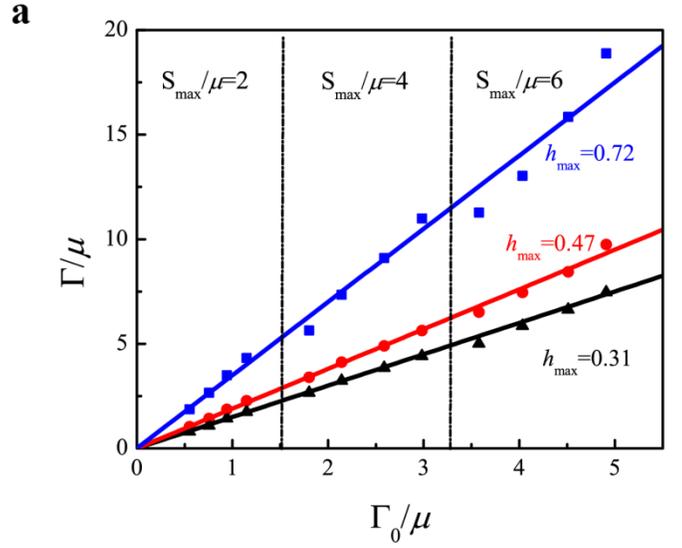

**b**

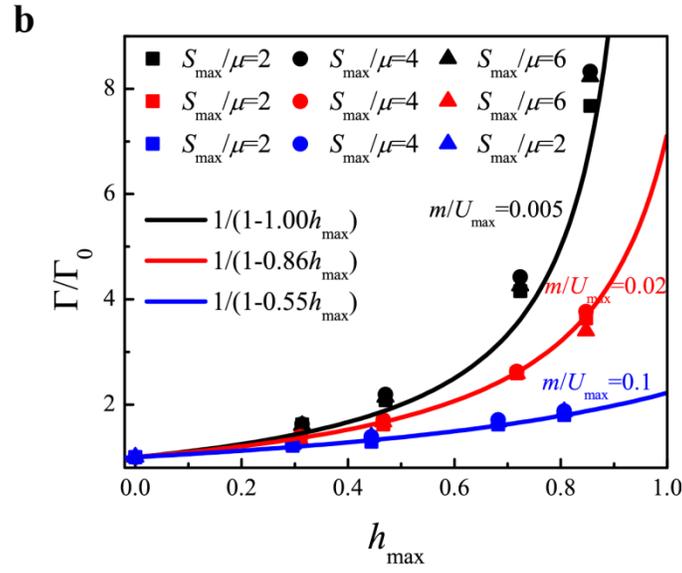

**c**

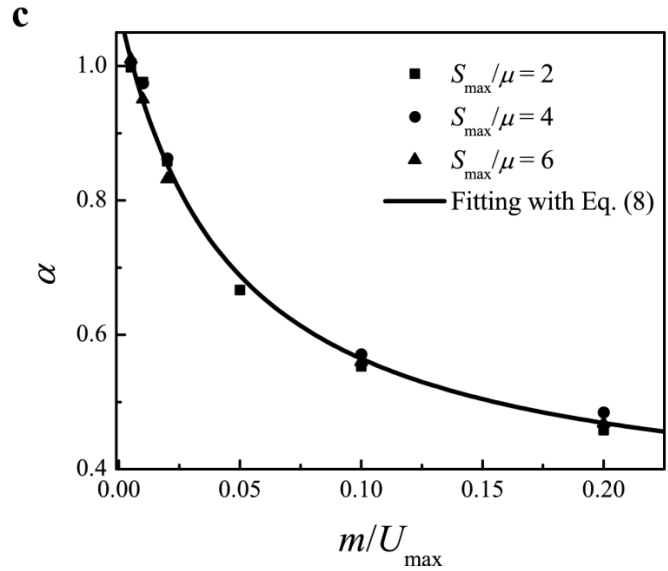

**Figure 3. Calculated fracture energies of soft materials from the coupled cohesive-zone and Mullins-effect model**. (a) Calculated $\Gamma$ as a function of $\Gamma_0$ for soft materials with different $S_{max}/\mu$ and $h_{max}$. The value of $m/U_{max}$ is set to be $0.01$. (b) Calculated $\Gamma$ as a function of $h_{max}$ for soft materials with various values of $S_{max}/\mu$ and $m/U_{max}$. (c) Calculated parameter $\alpha$ in Eq. (3) as a function of $m/U_{max}$.

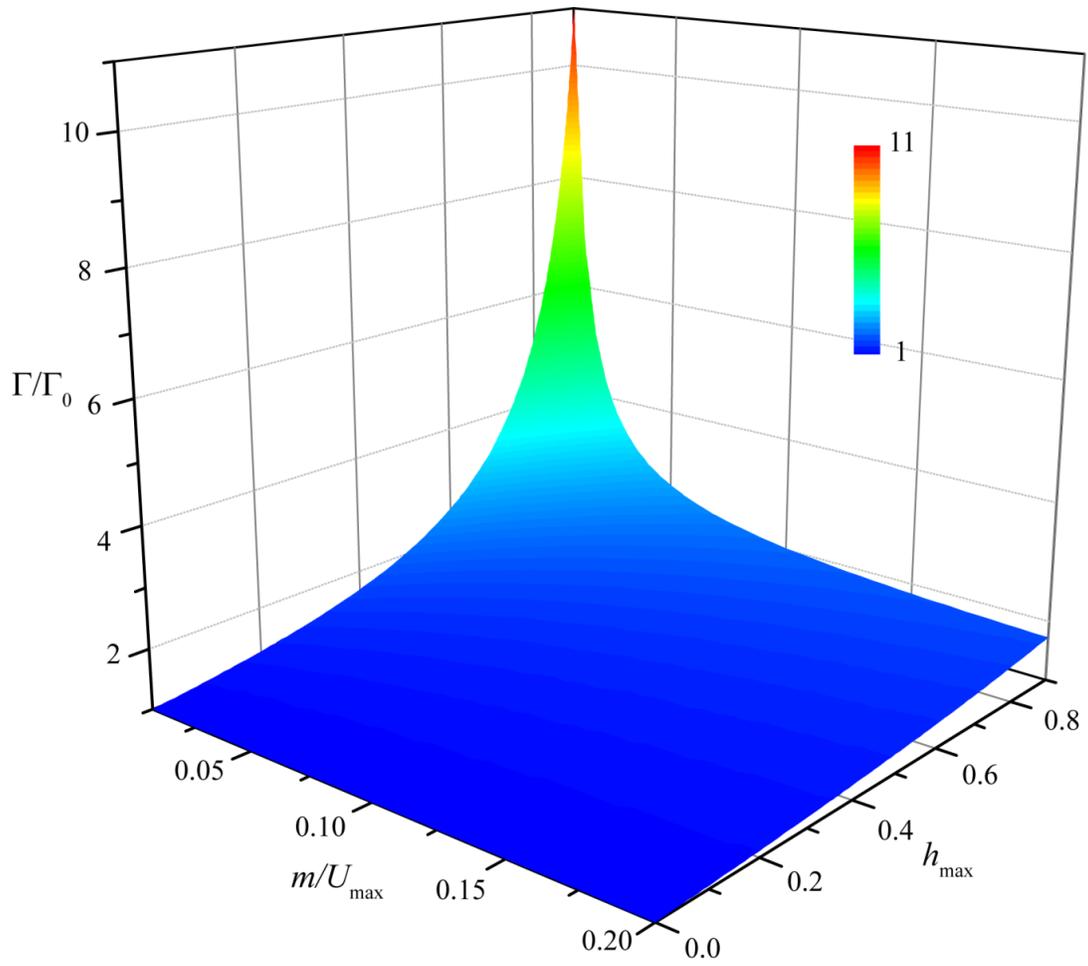

**Figure 4. A quantitative diagram for toughening mechanisms of soft materials**. The toughness enhancement ratio $\Gamma/\Gamma_0$ as a function of $h_{max}$ and $m/U_{max}$, calculated based on Eq. (3) and (8).

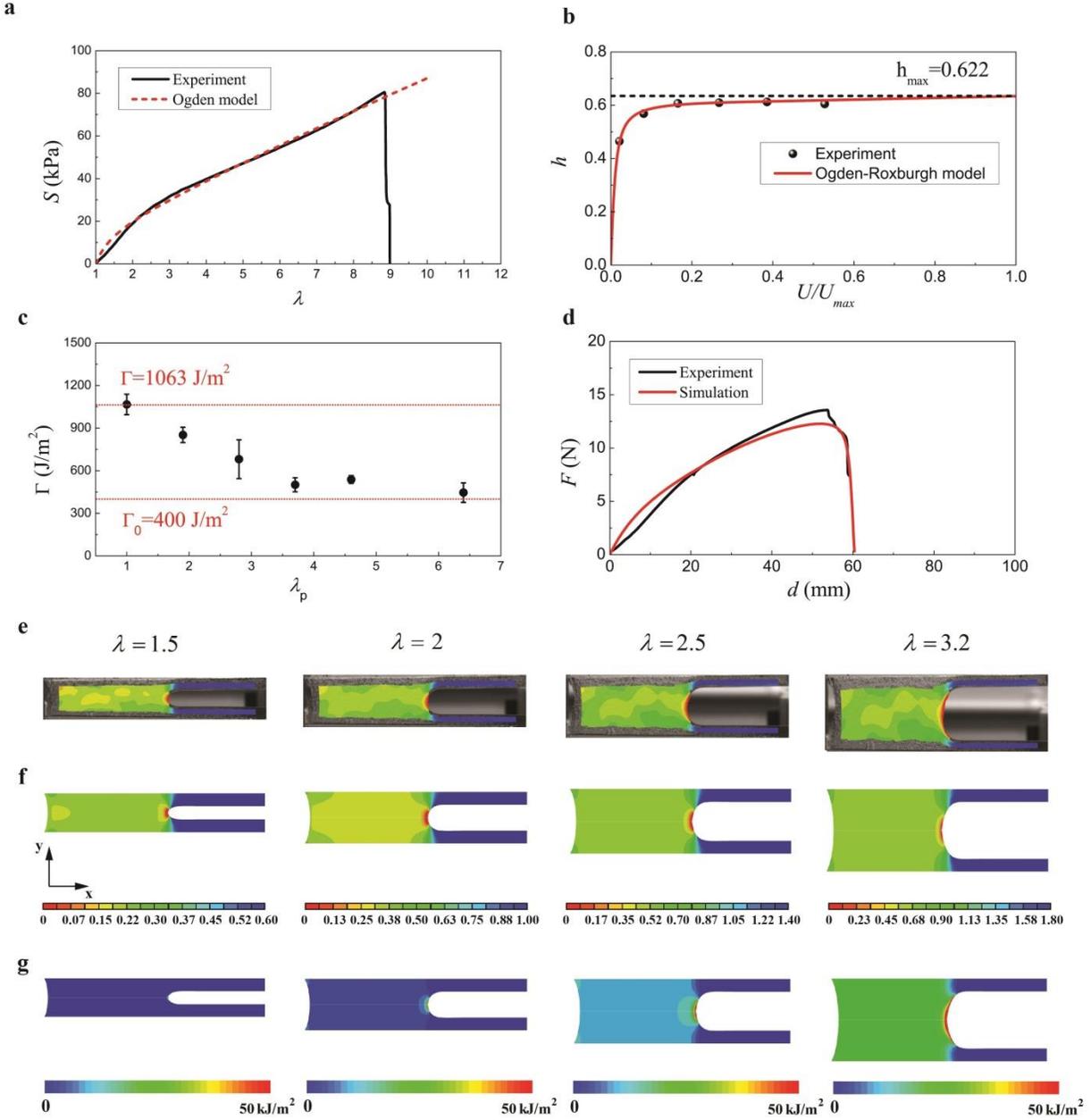

**Figure 5. Comparison between experiments and simulations on fracture of PAAm-alginate hydrogel Sample 1.** (a) Stress-stretch curve of the sample under pure-shear deformation and one term Ogden model, i.e., $\widetilde{W} = 2\mu / \alpha_1^2 \left( \lambda_1^{\alpha_1} + \lambda_2^{\alpha_1} + \lambda_3^{\alpha_1} - 3 \right)$, with $\mu = 10.81 \ kPa$ and $\alpha_1 = 1.879$. (b) The measured hysteresis ratios of the material deformed to different stretches, and the calculated hysteresis ratio by the modified Ogden-Roxburgh model with $r$=1.516 and $m$=4.274

J/m$^3$. (c) Measured fracture energy of the sample pre-deformed to different pre-stretches $\lambda_p$. (d) Force-displacement curves of the notched sample under pure-shear test measured from the experiment and predicted by the model. The strain field in the notched sample under pure-shear deformation to different stretches: (e) measured by DIC in the experiment and (f) predicted by the model. (g) Energy dissipated in the notched sample under pure-shear deformation to different stretches. The color represents the true strain ($\varepsilon_{yy}$) in (e) and (f) and the density of the energy dissipation in (g).